# Pressure dependence of the interfacial polarization and negative activation volume for dielectric relaxation in heterogeneous systems

Anthony N. Papathanassiou *

*National and Kapodistrian University of Athens, Department of Physics,*
*Section of Condensed Matter Physics,*
*Panepistimioplois, 15784 Zografos, Athens, Greece*

**ABSTRACT**

Negative activation volumes for dielectric relaxation are rarely reported in solid state physics and are limited to atomic scale processes. Broadband Dielectric Spectroscopy on heterogeneous systems, such as polycrystalline calcite and magnesite, hosting water in their porous space, detected intense dielectric relaxation mechanisms related to interfacial polarization. The characteristic relaxation frequency increased upon hydrostatic compression, indicating that the activation volumes for relaxation are negative. However, a theoretical interpretation for the negative sign of the activation volume is lacking up to date. Within the frame of effective medium approximation for mesoscopic heterogeneous two phase solid – fluid systems, we investigate how the synergy of the pressure dependencies of polarization and electric charge transport, respectively, dictate the pressure dependence of interfacial relaxation, predicting the value of the effective activation volume. Our theoretical approach succeeds in predicting the negative sign and magnitudes of the activation volume in water saturated polycrystalline materials.

(*) Corresponding author; email address: antpapa@phys.uoa.gr

I. INTRODUCTION

The electric and dielectric mismatch among different phases constituting a mesoscopic heterogeneous system, induce interfacial polarization effects. The characteristic relaxation frequency has long been well described by dielectric effective medium approximations. The relaxation of interfacial polarization in solid – water two phase heterogeneous materials, such as polycrystalline calcites and magnesite hosting water in their porous space was studied at various pressure and temperature states by employing Broadband Dielectric Spectroscopy during the past fourteen years (Papathanassiou, 2010; Papathanassiou, 2010; Papathanassiou, 2012; Papathanassiou and Sakellis, 2022). The analysis of the experimental results provided negative activation volumes for the interfacial polarization. Negative activation volumes for dielectric relaxation are rarely reported in solid state physics and refer to *atomic* scale relaxation, involving the rotation of permanent point defect dipoles or proton polaron hopping (Fontanella, 1982; Fontanella, 1997; Papathanassiou 2006; Papathanassiou, 2007). Until now, the question of the sign and magnitude of the activation volume for *interfacial* relaxation has not been addressed theoretically. The scope of the present work is to explain *why* activation volumes are negative and predict its magnitude, so as to compare with published experimental results. Our effort concerns pure condensed matter theory of the dielectric properties of heterogeneous systems and applied physics; laboratory experiments on solid – water two phase systems emerged from the need for gaining information about the relaxation phenomena in depth of the Earth's interior, related to the emission of transient electric currents preceding earthquakes, or, to interpret the frequency spectrum obtained from large scale induced polarization exploration for subsurface.

II. THEORETICAL BACKGROUND

The relaxation of an electric charge carrier from an equilibrium state to another one by overcoming a potential barrier, is characterized by the Gibbs free energy for relaxation $g^{act}$, defined as the difference of the Gibbs free energy $G^s$ of the material when a relaxing charge visits the inflection point (saddle point) of the potential barrier and the Gibbs free energy of the material $G^0$ when the charge is in its equilibrium ground state; i.e., $\boldsymbol{g^{act} = G^s - G^0}$ .

The transition rate Γ, at which a a charge surrounds the potential barrier successfully is given by (Varotsos, 1986) :

$$\Gamma = A\, \nu\, exp(-g^{act}/kT) \quad (1)$$

where $A$ is a geometrical factor, $\nu$ is an attempt frequency roughly equal to the frequency of the phonon assisting hopping and k is the Boltzmann's constant and T is the absolute temperature. The relaxation time $\tau$ is the mean time duration for an individual charge flow event, which equals the inverse of the transition rate (i.e., $\Gamma^{-1}$). Subsequently, taking the logarithm of eq. (1) and differentiating with respect to pressure, we get: $\left(\frac{\partial ln\tau}{\partial P}\right)_T = -\left(\frac{\partial ln\nu}{\partial P}\right)_T + \frac{1}{kT}\left(\frac{\partial g^{act}}{\partial P}\right)_T$. However, $\left(\frac{\partial g^{act}}{\partial P}\right)_T \equiv \boldsymbol{v^{act}}$, **where $\boldsymbol{v^{act}}$ denotes the activation volume and** $\left(\frac{\partial ln\nu}{\partial P}\right)_T = \gamma\kappa$, where $\gamma$ and $\kappa$ denote the Grüneisen parameter and the isothermal compressibility, respectively. Accordingly, $\boldsymbol{v^{act}}$ links to the percentage variation of the relaxation time $\tau$, at constant temperature (Varotsos, 1980; Varotsos, 1986):

$$v^{act} = kT\left\{\left(\frac{\partial ln\tau}{\partial P}\right)_T + \boldsymbol{\gamma\kappa}\right\} \quad (2)$$

The term $\gamma\kappa$ Is usually negligible compared to $\left(\frac{\partial ln\tau}{\partial P}\right)_T$. Hence, $v^{act}$ is a measure of the percentage change $\tau$ upon hydrostatic compression. Depending on the pressure evolution of the relaxation time, the activation volume can be positive, null or negative. Negative activation volumes were found in rare earth doped lead fluoride crystals (Fontanella, 1982) Nafion hydrogels (Fontanella, 1997) and, conducting polypyrrole (Papathanassiou 2006; Papathanassiou, 2007), as well as, water saturated polycrystalline minerals: Dielectric Spectroscopy has been used to study the dielectric relaxation in hydrated leukolite (polycrystalline magnesite) (Papathanassiou, 2010; Papathanassiou, 2012; Papathanassiou and Sakellis, 2022), at various limestone (Papathanassiou, 2010; Papathanassiou, 2012; Sakellis, 2014) and granodiorite (Sakellis, 2011; Papathanassiou, 2012). In Figs. 1 and 2, isotherms of the tangent of the dielectric loss angle $tan\delta$ as a function of frequency $f$ of hydrated polycrystalline magnesite (leukolite), reproduced from: (Papathanassiou 2010) and (Papathanassiou 2022), with permission from Elsevier, respectively. are presented. The peak maximum $f_{max,tan\delta}$ (which is proportional to the characteristic relaxation frequency $f_o$) shifts towards higher frequencies, thus, evidencing for negative activation volume. Experimental dielectric spectra (reproduced from: (Sakellis 2014) with permission from Elsevier for wetted limestone are depicted in Fig. 3. The observed dielectric relaxation mechanisms originate from the bulk material and not from electrode polarization (due to a double layer formation along the specimen-electrode interface) since the peak position, shape and intensity remained unaffected either on reducing the sample thickness, or modifying the nature of the electrodes attached on the parallel surfaces of specimen (silver or copper paste).

Hydrated rocks constitute heterogeneous systems consisting of solid matrix, water and, if partially saturated, air; the pressure variation of the relaxation time characterizing the transition among interfacial polarization states. An emerging question is **why** $v^{act}$ for relaxation is negative in the aforementioned hydrated rocks; i.e., which are the pressure dependencies of individual microscopic or mesoscopic processes that yield a detectable negative $v^{act}$. While there exist different models predicting $v^{act}$ for atomic or electron diffusion or drift (Flynn, 1968; Varotsos, 1986; Dyre, 2006): a model describing the pressure evolution (and, subsequently, the sign and value of $v^{act}$) of interfacial polarization and relaxation phenomena in mesoscopic heterogeneous systems is lacking. Beyond its importance from a theoretical view point, the detection and interpretation of $v^{act}$ for relaxation in fluid saturated rocks is important for different reasons, such as:

(a) Seismic Electric Signals (SES) emitted prior to earthquake occurrence, were interpreted as pressure induced polarization currents emerging from rocks, the relaxation of which are characterized by negative $v^{act}$; i.e., its characteristic relaxation time becomes shorter upon pressure (Varotsos *et al*, 1982; Varotsos and Alexopoulos, 1984; Varotsos, 2005).

(b) The spectral analysis of induced polarization used to profile fluids or gases confined in the porosity of sub-surface rocks (Revil, 2017; Qi, 2018), are influence by the temperature an pressure gradients developed from the Earth's surface towards its interior. Broadband dielectric data require reduction to elevated temperature and pressure states, so as to gain information at various depths from the Earth's surface. Accordingly, the activation energy $E^{act}$ and activation volume $v^{act}$ for relaxation are desirable from laboratory experiments and (Papathanassiou, 2012).

III. EFFECT OF PRESSURE ON RELAXATION IN TWO PHASE HETEROGENEOUS SYSTEMS

A porous dielectric, such as polycrystalline minerals, fully saturated with water, constitutes a two phase system. The minerals (solid phase) are commonly regarded as a dielectric insulator (the electrical conductivity is usually less than $10^{-10}$ S/cm (Chen, 2006)), while, water (liquid phase) is more conductive than the mineral. In the present work, we treat interfacial polarization due to electric mismatch of the end components; the electric double layer formation along both sides of the boundaries separating different phases is not discussed here. The dielectric properties of two phase heterogeneous materials and interfacial effects are treated within a differential effective medium approximation (DEMA) (Bruggeman, 1935; Sen *et al.*, 1981; Norris, 1985; Endres, 2000; Cosenza et al., 2003, Misra 2016), which complies with the experimental results. The main concept in effective medium approximations is to regard a heterogeneous mesoscopic system as an an effective homogeneous one. For a two-phase mixture, the DEMA designates one phase as a background matrix to which the other phase is added incrementally as inclusions (Hanai, 1968; Norris et al., 1985; Chelidze and Gueguen, 1999).

Assume a two phase system consisting of a solid mineral and water, respectively. We denote $\sigma_s$ and $\sigma_w$ the electrical conductivity of the solid mineral and water, respectively ($\sigma_w \gg \sigma_s$), $\varepsilon_s$ and $\varepsilon_w$ the relative (to the permittivity of free space $\varepsilon_o$) dielectric constant (of the solid mineral and water, respectively. We consider two configurations of a rock fully saturated with water:

a) The solid network is formed by aggregates of mineral grains and the inter-grain space is filled with water. Depending on the porosity and the microstructure, each grain (or agglomerates) are, more or less, are wetted by water.
b) The solid phase constitutes a porous dielectric matrix, whereas individual sparsely distributed pores are filled with water (Maxwell – Wagner – Sillars (MWS) model) (Sillars 1937).

Interfacial polarization phenomena occur in a heterogeneous material, due to electrical mismatch at the interfaces separating different phases. The dielectric relaxation of interfacial phenomena is described by a characteristic relaxation time $\tau$, which is a function of pressure and temperature. The inverse of the relaxation time is defined as the relaxation frequency:

$$f_0 \equiv \tau^{-1}. \quad (3)$$

Eqs (2) and (3) yield:

$$v^{act} = kT\left\{-\left(\frac{\partial \ln f_0}{\partial P}\right)_T + \boldsymbol{\gamma}\,\boldsymbol{\kappa}\right\} \quad (4)$$

**Model (a):** We assume that the mineral matrix consists of individual grains (or aggregates and the porosity and pore network that the water filling it, surrounds each solid aggregate (Chen 2006). Under these circumstances, the system can ba approximated as a mixture with solid electrically insulating embedded in water. According to Wagner's theory (van Beek, 1967) the relaxation frequency is (Chen, 2006):

$$f_0 = \frac{1}{2\pi\varepsilon_0} \frac{(2+\varphi_S)\sigma_W}{[2\varepsilon_W + \varepsilon_S + \varphi_S(\varepsilon_W + \varepsilon_S)]} \tag{5}$$

where $\varepsilon_0$ is the dielectric permittivity of free space, $\phi_s$ is the volume fraction of the solid inclusions. The isothermal pressure derivative of $lnf_0$ is:

$$\left(\frac{\partial lnf_0}{\partial P}\right)_T = \left(\frac{\partial \ln(2+\varphi_S)}{\partial P}\right)_T + \left(\frac{\partial ln\sigma_w}{\partial P}\right)_T - X \tag{6}$$

where:

$$X \equiv \frac{(\varepsilon_W + \varepsilon_S)\left(\frac{\partial \varphi_S}{\partial P}\right)_T + (2+\varphi_S)\left(\frac{\partial \varepsilon_W}{\partial P}\right)_T + (1+\varphi_S)\left(\frac{\partial \varepsilon_S}{\partial P}\right)_T}{2\varepsilon_W + \varepsilon_S + \varphi_S(\varepsilon_W + \varepsilon_S)} \tag{7}$$

- ***The pressure dependence of the volumetric fraction $\varphi_S$ of the solid phase.***

The volume fraction of the solid inclusions is defined as: $\varphi_S \equiv V_S/V$, where $V_s$ and $V$ denote the solid and total volumes, respectively. Let $V_W$ be the volume of the liquid phase. The isothermal compressibility of the mineral, water and fully saturated rock are: $\kappa_S = \left(-\frac{\partial lnV_S}{\partial P}\right)_T$, $\kappa_W = \left(-\frac{\partial lnV_W}{\partial P}\right)_T$ and $\kappa = \left(-\frac{\partial lnV}{\partial P}\right)_T$, respectively. The percentage variation of the volume fraction of the solid phase upon the confined pressure change is:

$$\left(\frac{\partial ln\varphi_S}{\partial P}\right)_T = \left(\frac{\partial lnV_S}{\partial P}\right)_T - \left(\frac{\partial lnV}{\partial P}\right)_T = -\kappa_S + \kappa \tag{8}$$

The compressibility $\kappa$ of the water saturated rock is a complex function of the porosity and its topology, which prevents us from knowing the isothermal compressibility $\kappa$ of the water saturated rock. Nevertheless, we can have a rough estimate of the value of the term $\left(\frac{\partial ln\varphi_S}{\partial P}\right)_T$: For hydrated limestone and hydrated magnesite, the isothermal compressibility is *0.052 GPa$^{-1}$* and *0.013 GPa⁻¹*, respectively (Wang, 1974). The isothermal compressibility of the corresponding minerals $\kappa_S$ are: *0.015 GPa⁻¹* for calcite and *0.013 GPa⁻¹* for magnesite (Papathanassiou, 1997). Subsequently, $\left(\frac{\partial \ln\_s}{\partial P}\right)_T = -\kappa_S + \kappa$ is estimated to be $0.037\ GPa^{-1}$ .for calcite grains surrounded by water and zero for magnesite inclusions in water.

- ***The pressure dependence of the electric charge transport and polarizabilities of the two end phases.***

The competition between the pressure derivatives of the reduced electrical conductivity and reduced dielectric constants of the individual materials composing the two phase system, determine the value (and the sign) of the pressure derivative of the characteristic relaxation frequency. The electrical conductivity of water at various temperatures and pressures have been published a long time ago by (Holtzpfel, 1969). Recent and more accurate measurements measurements were reported by (Liu, 2016); The low pressure data (i.e., in the limit of ambient pressure, at room temperature (308 K) yield $\left(\frac{\partial \ln \sigma_W}{\partial P}\right)_T \cong$ +9.2(9) $GPa^{-1}$. The (relative) dielectric constant and its pressure derivative are: $\varepsilon_W = 80$ and $\left(\frac{\partial \varepsilon_W}{\partial P}\right)_T = +33.0\ GPa^{-1}$ (Lide, 1994), respectively. For calcite mineral, at ambient conditions, $\varepsilon_s$ =8.45 (which is the mean of the values 8.2 to 8.7 cited in ref. (Shelby, 1980), respectively), and $\left(\frac{\partial \varepsilon_S}{\partial P}\right)_T = +0.11\ GPa^{-1}$ (Link, 1980). The dielectric constant of magnesite is 6.61 (Church 1988), but its pressure derivative has not been studied yet. For magnesite, we are therefore assuming that $\left(\frac{\partial \varepsilon_S}{\partial P}\right)_T$ is comparable to that of calcite.

The values of each term appearing in eq. (6) are depicted in Table I, for the for calcite and magnesite grains surrounded by water, respectively. We observe that, within the assumptions of the theoretical modeling and the experimental accuracy of the physical quantities involved in eq.(6), $\left(\frac{\partial \ln f_0}{\partial P}\right)_T$ is predicted to be positive. Accordingly, the activation volumes for relaxation calculated by employing eq. (4) are predicted to be negative.

**Model (b):** Consider a porous rock of an insulating mineral matrix, hosting a small fraction of identical ellipsoid pores. If the porosity is fully saturated with water, the hydrated rock can be treated according to the Maxwell-Wagner-Sillars (MWS) theory (Sillars, 1937):. This model dominates as a simple, clear and quite accurate to explain interfacial polarization phenomena in mesoscopic scale for about 90 years. Accordingly, a small fraction of conducting spheroid inclusions (of water) are distribute sparcely in a dielectric insulator of null electrical conductivity (Sillars 1937). The relaxation frequency $f_0$ of the interfacial polarization process, in the limit of small volume fractions of the inclusions, is (Sillars, 1937):

$$f_0 = \frac{1}{2\pi\varepsilon_0} \frac{\sigma_W}{[\varepsilon_W + (\lambda-1)\varepsilon_S]} \qquad (9)$$

where $\lambda$ $(\lambda>1)$ is a shape (and orientation) factor of an individual inclusion, which is assumed to be pressure and temperature invariant. $\lambda$ holds a unique value provided that the inclusions are identical and have the same axes oriented along the direction of the polarizing electric field. Note that, eq. (7) is – under the constrains asserted in the MWS model – independent upon the volumetric fraction, which, in turn, simplifies calculations.. By differentiating the natural logarithm of eq. (9) with respect to pressure, we get:

$$\left(\frac{\partial ln f_0}{\partial P}\right)_T = \left(\frac{\partial ln \sigma_w}{\partial P}\right)_T - Y \qquad (10)$$

where $Y$ is defined as:

$$Y \equiv \frac{\left(\frac{\partial \varepsilon_W}{\partial P}\right)_T + (\lambda-1)\left(\frac{\partial \varepsilon_S}{\partial P}\right)_T}{[\varepsilon_W + (\lambda-1)\varepsilon_S]} \qquad (11)$$

The terms composing eq. (10), the values of $\left(\frac{\partial ln f_0}{\partial P}\right)_T$ and, according to eq. (4), $\frac{v^{act}}{kT}$, for calcite or magnesite matrix containing a small fraction of sparcely dispersed spheroid voids, filled with water, are given in Table II.

IV. CONCLUSIONS

In conclusion, comparing the values enlisted in Table I and 2, respectively, we conclude that, while the models for water saturated carbonate rocks discussed within models (a) and (b) respectively, represent two extreme limits (i.e., solid grains in a background of water and a porous solid matrix hosting sparsely distributed water inclusions, respectively) converge in predicting negative valued activation volume for interfacial relaxation. The reason is that pressure affects significantly the electrical conductivity of water and, therefore, the value of $\left(\frac{\partial ln\sigma_w}{\partial P}\right)_T$ is much larger than the pressure derivatives of the dielectric constants of the end members, as well as, the pressure derivative of the volumetric fraction $\varphi_S$. $\left(\frac{\partial lnf_0}{\partial P}\right)_T$ is found to be positive , which, in turns - according to eq. (4) - yields negatively signed activation volume. Broadband Dielctric Spectroscopy experiments at various pressures yielded the following values for $\frac{v^{act}}{kT}$ : . $-20\ (4) GPa^{-1}$ for hydrated leukolite (polycrystalline magnesite) (Papathanassiou, 2010), $-4.3\ (4) GPa^{-1}$ limestone (Papathanassiou, 2010; Papathanassiou, 2022) and $-5.9\ (7)\ GPa^{-1}$ for hydrated limestone (Sakellis, 2014). The models treated in the present work to simulate interfacial relaxation in rocks saturated with water succeeds in predicting a negative sign for $v^{act}$. As can be seen in Tables 1 and 2, the predicted $v^{act}/kT$ values are of the same order of magnitude compared with the experimental ones. Among the two effective medium approximations treated in the present work, i.e., a dilute limit approximation for $\varphi_S \to 0$ and MWS one, respectively, the second one may apply for the detection of pressure induced polarisation currents emerging from rocks or the profiling of fluids in the pores of sub-surface rocks). The second model is used to show that the pressure dependence of the relaxation time is does significantly depend on the microstructure. We point out that eqs. (6) and (10) for the pressure dependence of the characteristic relaxation frequency $f_0$ and $v^{act}/kT$ is based on well established dielectric effective medium approximations, which, oversimplify the porous network of the rock, ignore its topological and distribution characteristics and ignore thee electric double layer formation in the vicinity of interfaces. Within this frame of assumptions and simplifications, the predicted $v^{act}/kT$ values are of the same sign and same order of magnitude of those determined experimentally.

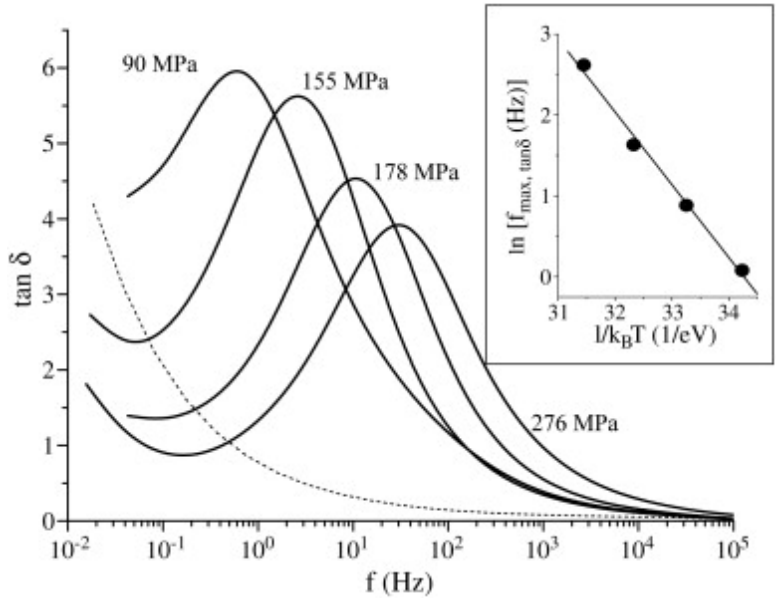

Fig. 1. Isotherms (T=349K) of the tangent of the dielectric loss angle $tan\delta$ as a function of frequency $f$ of hydrated polycrystalline magnesite (leukolite). The dash line is a typical measurement on an as-received (not hydrated) specimen at 178 MPa. In the inset diagram, the natural logarithm of the peak maximum $f_{max,tan\delta}$ (which is proportional to the characteristic relaxation frequency $f_o$) is plotted against $1/kT$ for P=155MPa. The straight line fitted to the data points ensures thatEq. (1) is valid. Figure is reproduced from: (Papathanassiou 2010) with permission from Elsevier.

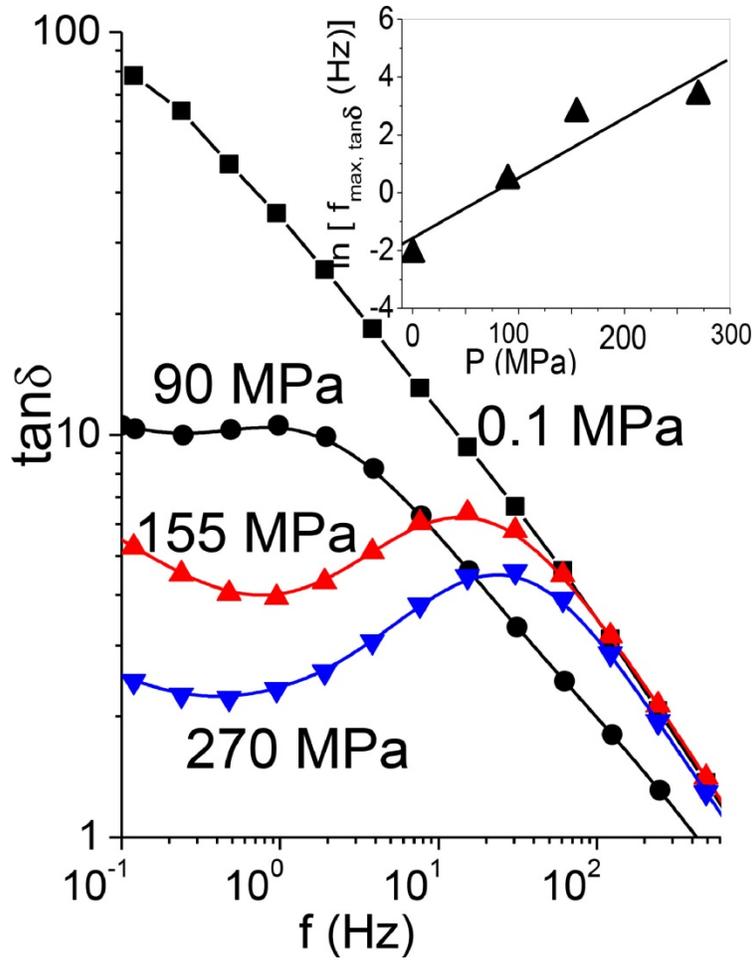

Fig. 2. Tangent loss angle spectra ofhydrated polycrystalline magnesite (leukolite) recorded at 370 K and different pressure values. Inset: the natural logarithm of the maximum (resonance) frequency as a function of pressure; the straight line that fits the data points, As explained in the text, its positive slope yields a negative activation volume for relaxation. Figure is reproduced from: (Papathanassiou 2022) with permission from Elsevier.

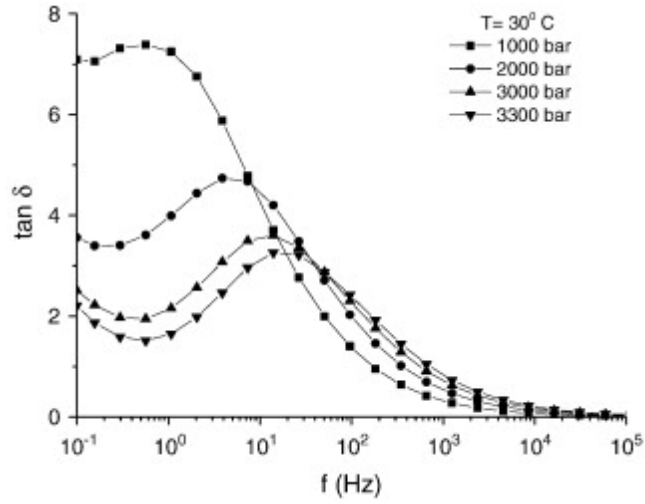

Fig. 3. Isotherms (T=303 K) of *tanδ* vs frequency of wetted limestone at various pressures. The peak maximum $f_{max,tan\delta}$ (which is proportional to the characteristic relaxation frequency $f_o$) shifts towards hogher frequencies on compression, yielding a negative activation volume. Figure is reproduced from: (Sakellis (2014) with permission from Elsevier).

## Table I.

*The values of the pressure derivatives of $\varphi_S$ and $ln\sigma_W$, as well as, of the term X, which were estimated from published experimental works (details and citations are given in the text), that lead to the theoretical pressure derivative of $lnf_0$, via eq. (6), for solid grains in a background of water discussed ithin the frame of model (a). Since X is a function of $\varphi_S$, its minimum and maximum values are presented.. Term $\gamma\kappa$ appearing in eq. (4) was taken from (Papathanassiou, 2010) for hydrated specimens, to calculate $v^{act}/kT$. All quantities are in $GPa^{-1}$ units. We assume that the main source of error in calculating $\left(\frac{\partial lnf_0}{\partial P}\right)_T$ and $v^{act}/kT$ is that in the pressure derivative of the electrical conductivity of water and (since the accuracy of other quantities are not mentioned im the literature) should be considered as minimum uncertainties.*

| Solid phase | Eq. (6) | | | Eq (4) | | | Experimental |
|---|---|---|---|---|---|---|---|
| | $\left(\frac{\partial ln\varphi_S}{\partial P}\right)_T$ | $\left(\frac{\partial ln\sigma_w}{\partial P}\right)_T$ | X | $\left(\frac{\partial lnf_0}{\partial P}\right)_T$ | $\Gamma\kappa$ | $\frac{v^{act}}{kT}$ | $\frac{v^{act}}{kT}$ |
| Calcite | 0.037 | 9.2(9) | 0.41…0.59 | 8.6(9)…8.8(9) | 0.022 | -8.6(9)… − 8.8(9) | −4.3 (4)[a] −5.9 (7)[b] |
| Magnesite | ~ 0 | 9.2(9) | 0.41…0.58 | -8.1(9)…8.8(9) | 0.088 [a] | -8.1(9)… − 8(9) | −20 (4)[a,c] |

[a] Ref. (Papathanassiou, 2010)
[b] Ref. (Sakellis, 2014).
[c] Ref. (Papathanassiou, 2022).

## Table II.

*The values of the pressure derivative of $ln\sigma_W$, as well as, of the term Y, which were estimated from published experimental works (details and citations are given in the text), that lead to the theoretical pressure derivative of $lnf_0$, via eq. (8), for a solid matrix containing a very small fraction of water inclusions discussed within model (b). The term $\gamma\kappa$ appearing in eq. (4) was taken from (Papathanassiou, 2010) for hydrated specimens, to calculate $v^{act}/kT$. All quantities are in $GPa^{-1}$ units.*

| Solid phase | Eq. (10) | | Eq (4) | | | Experimental |
|---|---|---|---|---|---|---|
| | $\left(\frac{\partial ln\sigma_w}{\partial P}\right)_T$ | Y | $\left(\frac{\partial lnf_0}{\partial P}\right)_T$ | $\gamma\kappa$ | $\frac{v^{act}}{kT}$ | $\frac{v^{act}}{kT}$ |
| Calcite | 9.2(9) | 0.34 | 8.9(9) | 0.022 | -8.9(9) | −4.3 (4)[a] −5.9 (7)[b] |
| Magnesite | 9.2(9) | 0.36 | 8.8(9) | 0.088 [a] | -8.8(9) | −20 (4)[a,c] |

[a] Ref. (Papathanassiou, 2010).
[b] Ref. (Sakellis, 2014).
[c] Ref. (Papathanassiou, 2022).

## DATA AVAILABILITY STATEMENT

No new data were used in the present work.

# REFERENCES

Bruggeman, D. A. G. (1935), Berechnung verschiedemer physikalischer Konstanten von hetarogenen Substanzen, Ann. Phys., **24**, 636–679.

Chelidze, T. L., and Gueguen Y. (1999), Electrical spectroscopy of porous rocks: A review—I. Theoretical models, Geophys. J. Int., **137**, 1–15.

Chen, Y., and Or, D. (2006) Geometrical factors and interfacial processes affecting complex dielectric permittivity of partially saturated porous media, Water Resour. Res., **42**, W06423, doi:10.1029/2005WR004744.

Church J., R. H., Webb W. E., and Salsman, J. B. (1988) Dielectric Properties of Low-Loss Minerals, RI 9194 Report of Investigations. (https://stacks.cdc.gov/view/cdc/10143/cdc_10143_DS1.pdf)

Cosenza, P., C. Camerlynck, and Tabbagh, A. (2003) Differential effective medium schemes for investigating the relationship between high-frequency relative dielectric permittivity and water content of soils, Water Resour. Res., **39**, 1230, doi:10.1029/2002WR001774.

Dyre, J. C. (2006) The glass transition and elastic models of glass-forming liquids, Rev.Mod.Phys., **78**, 953—972.

Endres, A. L., and J. D. Redman (1996) Modeling the electrical properties of porous rocks and soils containing immiscible contaminants, J. Environ. Eng. Geophys., **1**, 105–112.

Flynn, C. P.< (1968) Atomic migration in monoatomic crystals, Phys, Rev., **171**, 682 – 698.

Fontanella, J., Wintersgill M.C., Figueroa D.R., Chadwick A.V. and Andeen C.C. (1982) Anomalous pressure dependence of dipolar relaxation in rare earth doped lead fluoride, Phys. Rev. Lett., **51**, 1892-1895.

Fontanella, J.J., Edmondson, C.A., Wintersgill, M.C., Wu, Y., Greenbaum, S.G. (1996) High-pressure electrical conductivity and NMR studies in variable equivalent weight NAFION membranes. Macromolecules, **29**, 4944–4951. doi:10.1021/ma9600926.

Hanai, T. (1968), Electrical properties of emulsions, in Emulsion Science, edited by P. Sherman, pp. 354–478, Elsevier, New York.

Lide, D. R., (1994) CRC Handbook of Chemistry and Physics; CRC press, 74$^{th}$ ed., Table 6-15.

Link, J., Wintersgill, M. C., Fontanella, J. J., Bean, V. E., Andeen; C. G. (1981) Pressure variation of the low-frequency dielectric constants of some anisotropic crystals. J. Appl. Phys., **52**,936–939. https://doi.org/10.1063/1.328780.

Liu, B., Gao, Y., Han, Y., Ma, Y.,, Gao, C. (2016) In situ electrical conductivity measurements of $H_2O$ under static pressure up to 28 GPa, Physics Letters A, **380**, 2979-2983,. https://doi.org/10.1016/j.physleta.2016.07.007.

Misra, S.,Torres-Verdín, C., Revil, A., Rasmus, J.,Homan, D. (2016) Interfacial polarization of disseminated conductive minerals in absence of redox-active species — Part 1: Mechanistic model and validation,Geophysics, **81**, E139 – E157,

Norris, A. N., P. Sheng, and A. J. Callegari (1985) Effective-medium theories for two-phase dielectric media, J. Appl. Phys., **57**, 1990–1996.

Holtzpfel, W. B. (1969) Effect of Pressure and Temperature on the Conductivity and Ionic of Water up to 100 kbar and 1000°C, J. Chem. Phys., **50**, 4424-8.

Papathanassiou, A. N. (1997) cntribution of the pressure variation of the porsity to the activation volume evaluation from ionic conductivitymeasurements under pressure, J. Phys. Chem. Solids, **58**, 2107 – 2111.

Papathanassiou, A. N. (2001) Pressure Variation of the Conductivityin Single Crystal Calcite, phys. stat. sol. (b), **228**, , R6–R7.

Papathanassiou, A.N., Sakellis, I., Grammatikakis, J. (2006) Separation of electric charge flow mechanisms in conducting polymer networks under hydrostatic pressure. Appl. Phys. Lett., **89**, 222905. doi:10.1063/1.2768623.

Papathanassiou, A.N., Sakellis, I., Grammatikakis, J. (2007) Migration volume for polaron dielectric relaxation in disordered materials. Appl. Phys. Lett., **91**, 202103. doi:10.1063/1.2812538).

Papathanassiou, A. N., Sakellis, E., Grammatikakis, J. (2010) Negative activation volume for dielectric relaxation in hydrated rocks, Tectonophysics, **490**, 307-309. doi:10.1016/j.tecto.2010.04.030

Papathanassiou, A. N., Sakellis, E., Grammatikakis, J. (2012) Dielectric relaxation under pressure in granular dielectrics containing water: Compensation rule for the activation parameters, Solid State Ionics, **1-4**, 209 – 210. http://dx.doi.org/10.1016/j.ssi.2011.12.011

Papathanassiou, A. N., Sakellis, E. (2022) Pressure and temperature dependence of the relaxation of the electrical, double layer in hydrated magnesite rock (leukolite), Results in Geophysical Sciences, **11**, 100044. https://doi.org/10.1016/j.ringps.2022.100044

Qi, Y., Soueid, A., Revil, A., Ghorbani, A., Abdulsamad, F., Florsch, N. (2018) Induced polarization response of porous media with metallic particles: Part 7: Detection and quantification of buried slag heaps, Geophysics, **83**, E277–E291. 10.1190/GEO2017-0760.1

Revil, A., Deqiang Mao, I, , Zhenlu Shao, , Sleevi, M. F. (2017) Induced polarization response of porous media with metallic particles — Part 6: The case of metals and semimetals, Geophysics, **82**, E97-E110. DOI: 10.1190/GEO2016-0389.1

Sakellis, E., Papathanassiou, A. N., Grammatikais, J. (2011) Dielectric properties of granodiorite partially saturated with water and its correlation to the detection of seismic electric signals, Tectonophysics, **511**, 148 – 151. https://doi.org/10.1016/j.tecto.2011.09.012

Sakellis, E., Papathanassiou, A. N., Grammatikais, J. (2014) Measurements of the dielectric properties of limestone under pressure and their importance for seismic electric signals, J. Appl. Geohysics, **102**, 77 – 80, https://doi.org/10.1016/j.jappgeo.2013.12.013

Sen, P. N. (1981) Relation of certain geometrical features to the dielectric anomaly of rocks, Geophysics, **46**, 1714–1720.

Shelby, R., Fontanella, J., Andeen, C. (1980) The low temperature electrical properties of some anisotropic crystals, Journal of Physics and Chemistry of Solids, **41**, , 69-74 ttps://doi.org/10.1016/0022-3697(80)90122-5.

Sillars, R. W. (1937) The properties of a dielectric containing semiconducting particles of various shapes, J. Inst. Elec. Engrs, **80**, 378. 10.1049/jiee-1.1937.0058

van Beek, K. H., (1967) Progress in Dielectrics, edited by J. B. Birks Hey-wood, London, Vol. 7, p. 69.

Varotsos, P., Alexopoulos, K. (1980) On the question of the calculation of migration volume in ionic crystals, Phil. Magazine A, **42**, 13-18.

Varotsos, P., Alexopoulos, K. (1984) Physical properties of the variation of the electric field of the earth preceding earthquakes: determination of the epicenter and magnitude. Tectonophysics, **110**, 99–125.

Varotsos, P. A., Alexopoulos, K. (1986) Thermodynamics of Point Defects and Their Relation with Bulk Properties, North – Holland.

Varotsos, P., Alexopoulos, K., Nomicos, K. (1982) Comments on the physical variation of the Gibbs energy for bound and unbound defects. Phys. Stat. Sol. (b) **111**, 581–590.

Varotsos, P.A. (2005) The Physics of the Seismic Electric Signals, TERRAPUB, Tokyo

Wang Chi-Yuen (1974) Pressure coefficient of compressional wave velocity for a bronzite, J. Geophys. Res., **79**, 771-772.